\documentclass[Afour,sageh,times]{sagej}

\usepackage{moreverb,url}
\usepackage{amsmath}
\usepackage[colorlinks,bookmarksopen,bookmarksnumbered,citecolor=red,urlcolor=red]{hyperref}
\usepackage{cleveref}
\usepackage{endfloat}
\usepackage{color,soul}
\usepackage{graphicx}
\usepackage{subfig}
\usepackage{threeparttable}
\setcitestyle{super,open={},close={}}

\newcommand\BibTeX{{\rmfamily B\kern-.05em \textsc{i\kern-.025em b}\kern-.08em
T\kern-.1667em\lower.7ex\hbox{E}\kern-.125emX}}

\def\volumeyear{2016}

\begin{document}

\runninghead{Roychoudhury et al.}

\title{Beyond p-values: a phase II dual-criterion design with
  statistical significance and clinical relevance}

\author{Satrajit Roychoudhury\affilnum{1}, Nicolas
  Scheuer\affilnum{2}, and Beat Neuenschwander\affilnum{2} }

\affiliation{\affilnum{1} Pfizer Inc, USA\\
\affilnum{2} Novartis Pharma AG, Switzerland
}

\corrauth{Satrajit Roychoudhury, 
Pfizer Inc,
New York, NY, USA}

\email{satrajit.roychoudhury@pfizer.com}

\begin{abstract}
\vspace{5 mm}

Background
\\
Well-designed phase II trials must have acceptable error rates
relative to a pre-specified success criterion, usually a statistically
significant p-value. Such standard designs may not always suffice from
a clinical perspective because clinical relevance may call for more. For
example,  proof-of-concept in phase II often requires not only statistical
significance but also a sufficiently large effect estimate.  
\\   

Purpose
\\
We propose dual-criterion designs to complement statistical
significance with clinical relevance, discuss their methodology, 
and illustrate their implementation in phase II.
\\

Methods
\\
Clinical relevance requires the effect estimate to pass a clinically
motivated threshold (the decision value). In contrast to standard
designs, the required effect estimate is an explicit design input
whereas study power is implicit. The sample size for a dual-criterion
design needs careful considerations of the study's operating characteristics
(type-I error, power).
\\

Results
\\
Dual-criterion designs are discussed for a randomized controlled and a
single-arm phase II trial, including decision criteria, sample size
calculations, decisions under various data scenarios, and operating
characteristics. The designs facilitate GO/NO-GO decisions due to
their complementary statistical-clinical criterion.
\\

Limitations
\\
While conceptually simple, implementing a dual-criterion design needs
care. The clinical decision value must be elicited carefully in
collaboration with clinicians, and understanding similarities and
differences to a standard design is crucial.
\\

Conclusion
\\
To improve evidence-based decision-making, a formal yet transparent
quantitative framework is important. Dual-criterion designs offer an
appealing statistical-clinical compromise, which may be preferable to
standard designs if evidence against the null hypothesis alone
does not suffice for an efficacy claim.

\vspace{5 mm}

\end{abstract}


\maketitle

\section{Introduction}
Making evidence-based decisions is essential in drug development. To
conclude activity of a drug, such decisions usually require a
statistically significant p-value from a well-designed trial. However,
statistical significance alone may not suffice if the trial does not
deliver a clinically relevant effect estimate (Wasserstein and Lazar
\cite{Wasserstein2016}). We discuss a simple dual-criterion
design that formally combines a statistical (significance) and
clinical (effect size estimate) criterion.

Here we focus on phase II trials, which play a key role in drug
development. In particular, Proof-of-Concept (PoC) is a milestone, the
earliest point when the efficacy of an experimental drug is formally
investigated. PoC trials typically lead to one of the following
decisions: continue further development (GO), stop development
(NO-GO), or seek further information.

The consequences of false decisions in phase II can be
severe. Therefore, irrespective of the chosen design, understanding
statistical error rates is important. For example, false-positive
results (type-I errors) in PoC trials will lead to phase III trials
with ineffective experimental drugs and likely cause negative
results. On the other hand, false negative results (type-II errors)
will often stop the development of a potentially useful drug. Various
phase II design options have been proposed, with primary focus on
error rate control (e.g. Fleming \cite{Flemming1982}, Herson and Carter
\cite{Herson1986}, Simon \cite{Simon1989}, Schaid et
al. \cite{Schaid1990}, Storer \cite{Storer1992}, Liu et
al.\cite{Liu1993}, \cite{Liu1999}, Sargent et al. \cite{SARGENT2001},
Korn et al. \cite{Korn2001}, Rubinstein et al. \cite{Rubinstein2005},
Simon et al.\cite{Simon2005}, Parashar et al. \cite{Parashar2016}).

GO or NO-GO decisions should always be seen in the context of the
clinical needs and the competitive landscape (Cartwright et
al. \cite{cartwright2010}), which requires discussions among clinical
teams, governance boards, and key opinion leaders. Consensus is
important and easier to reach if trial results meet a pre-defined
clinical criterion. A few proposals to formally account for a
relevant effect size in clinical and other settings have been
discussed (Nicewander and Price \cite{nicewander1997}, Chuang-Stein et
al. \cite{chuang2011a},\cite{chuang2011b}, Neuenschwander et
al. \cite{neuenschwander2011}, Fisch et al. \cite{fisch2015}, Frewer
et al.\cite{frewer2016}). Here, we will discuss a simple
dual-criterion, which we have applied in many phase II
trials.

We will first review three generic phase II designs: the
standard, dual-criterion, and precision design. We then describe the
dual-criterion design in more detail and show a frequentist and
Bayesian application. 

\section{Design choices for  phase II trials }
In this section, we introduce the main components of three basic phase II designs:
the standard, dual-criterion, and precision design (Table \ref{SDCPdesign}). 

\subsection{The standard design}
The standard design is very popular in drug development. For
comparative (treatment vs. control) trials, it puts forward stringent
statistical criteria expressed as error rates:
\begin{enumerate}
\item \emph{type-I error control}: if there is no treatment
  effect (the \emph{null hypothesis}), the probability to declare success at the
  end of the trial is at most $\alpha$. Typical values for $\alpha$
  (one-sided) are 2.5\% for confirmatory trials, and 5\% or 10\% for
  early phase trials. Here, success means statistical significance.
\item \emph{power}: for a selected effect size of interest (the
  \emph{alternative hypothesis}), the probability of trial success is 80\% or 90\%.
\end{enumerate}
While the standard design prevails in the confirmatory setting, its
limitations are well-known (Senn \cite{senn1997}). The main issue is that
statistical significance only guarantees sufficient evidence to reject
the null hypothesis (no effect), which may not suffice from a clinical
perspective. Importantly, statistical
significance does not necessarily support the alternative hypothesis,
except when the p-value is highly significant.

\subsection{The dual-criterion  design}
The dual-criterion design addresses the limitation of the standard
design. It is appealing if one wants to base trial success not only on
statistical significance but also on the effect estimate, which is
more tangible for clinicians. The design requires the following inputs
(Table \ref{SDCPdesign}):
\begin{enumerate}
\item Like the standard design, type-I error control is needed, which
  requires a null hypothesis and type-I error $\alpha$. Eventually, a
  significant p-value will provide sufficient evidence for a positive
  treatment effect.
\item Unlike the standard design, the dual-criterion design requires a
  decision value, which is clinically motivated. The decision value is
  the minimum effect estimate needed for trial success. Estimates
  superior to this value justify a GO decision, whereas inferior
  values do not and will usually require additional considerations.
\end{enumerate}
Compared to the standard design, the dual-criterion design combines
the rigid statistical criteria (significant p-value, type-I error
control) with a clinical criterion that guarantees a
sufficiently large effect estimate.

\subsection{The precision design}
The two previous designs rely on error rates (type-I error, power),
which require clearly defined criteria for trial success. For the
special case where the null hypothesis and other benchmark values
(alternative hypothesis or decision value) cannot be determined, a
precision design may be an option. It requires a sufficiently precise
effect estimate, usually defined via the width of the 95\% confidence
interval.

\subsection{Example}
We now illustrate the three designs with a simple example.  Consider a
randomized trial with patients equally randomized to the experimental
drug and control. The outcome of interest is the hazard-ratio (HR) for
progression-free survival (PFS), where the HR is less than one if the
experimental drug is better than the control.
 
The standard design requires a null hypothesis with corresponding
type-I error and an alternative parameter with corresponding power
(Table \ref{threedesigns}). For example, assume a superiority trial
(null hypothesis HR=1), one-sided type-I error 2.5\%, and 90\% power
under a 25\% hazard reduction (alternative hypothesis HR=0.75). The
number of events to meet these requirements is 508. For this design,
statistical significance is achieved if the estimated HR is 0.84, a
hazard reduction of 16\%. Note that this estimate threshold is
implicit; that is, it follows from the design specifications.

The dual-criterion design requires a null hypothesis (with
corresponding one-sided type-I error $\alpha$) and the decision value
for the effect estimate. As seen above, success for the standard
design (with n=508 events) requires an estimated hazard reduction of
at least 16\%. If clinical considerations require an estimated
reduction of at least 20\%, the number of events must be at least 309
(see next section for details). Importantly, the dual-criterion design
also controls the type-I error at 2.5\%, but the power is implicit;
that is, it follows from the type-I error and the decision value
criterion. For 309 events, the power is smaller than for the standard
design. The reason is simple: an estimated hazard reduction of 20\%
(the clinical criterion) is more difficult to achieve than a 16\%
hazard reduction (considered clinically insufficient) for the standard
design.

For the precision design, which assumes no comparative benchmarks
(null hypothesis, alternative hypothesis, decision value), assume that the
aim is to estimate the HR with "20\% precision": 95\%-interval =
(HR/1.2, HR$\times$1.2). This requires 462 events. For "25\%
precision", 95\%-interval = (HR/1.25, HR$\times$1.25), the number of
events is 309.

\begin{table*}[ht]
\small\sf\centering
\caption{Design components for standard, dual-criterion, and
  precision designs.  
\label{SDCPdesign}}
\begin{tabular}{ p{4cm} p{4cm} p{4cm} p{4cm}}
     &  standard design	& dual-criterion design	& precision design \\[2mm]
null value & required & required &  - \\
alternative  value & required  & -  & - \\
decision value & (implied) & required & - \\
type-I error & required & required & - \\
power & required & (implied) & - \\
precision of estimate & (implied) & (implied) & required\\
sample size & (implied) & (implied) & (implied) \\
\end{tabular}
\label{threedesigns}
\end{table*}

\subsection{Dual-criterion designs: criteria, outcomes, and decisions}
The dual-criterion design is recommended for comparative trials if the
required evidence for an effect can be expressed for two benchmarks:
the null value (null hypothesis) and the decision value for the effect
estimate. For a GO-decision, the evidence must be sufficiently strong
that the effect is better than the null value and that the effect
estimate reaches the clinically relevant decision value.  The criteria are
\begin{enumerate}
\item \emph{statistical significance}: the one-sided p-value must be
  less than $\alpha$. Equivalently, the one-sided $100(1-\alpha)\%$
  confidence interval must exclude the null value.
\item	\emph{clinical relevance}: the estimated effect must reach the decision value.
\end{enumerate}
Regarding the statistical criterion, 10\%, 5\%, and 2.5\% are commonly
used for the one-sided type-I error $\alpha$, depending on the
required strength of evidence to reject the null hypothesis. The
clinical criterion must be elicited from clinical experts.  Table
\ref{dcdecision} shows the four possible outcomes of a dual-criterion
design:
\begin{itemize}
\item case 1: if trial results fail both criteria, the decision will
  be a NO-GO.
\item case 2: if the statistical and clinical criteria are both met,
  the decision will be a GO.
\item cases 3 and 4: inconclusive situations arise when only one
  criterion holds. For example, for large trials or unexpectedly small
  variability of the outcome, statistical significance may hold, yet
  the estimate may miss the decision value (case 3). On the other hand,
  for small trials or unexpectedly large variability of the outcome,
  the effect estimate may pass the decision value, but statistical
  significance may not hold (case 4).
\end{itemize}

\begin{table}[h]
\captionsetup{justification=centering}
\begin{minipage}[b]{0.5\linewidth}
\centering
\begin{tabular}{ | c | c | c |}
    \hline
                  & not relevant & relevant \\ 
 \hline
                           &  (1)  &  (4) \\
 not significant    &  \textbf{NO-GO}  & \textbf{ inconclusive} \\
                           & insufficient evidence  &  \\ 
                           & of relevant efficacy &  \\ \hline
                           &  (3)  &  (2) \\
 significant    &  \textbf{inconclusive}   &  \textbf{GO}  \\
                           &                       & evidence of \\
                           &                       & relevant efficacy \\ \hline
   \end{tabular}
\par\vspace{0pt}
\end{minipage}
\begin{minipage}[b]{0.5\linewidth}
\centering
\includegraphics[trim={0  16cm 0 0},clip, scale=0.5]{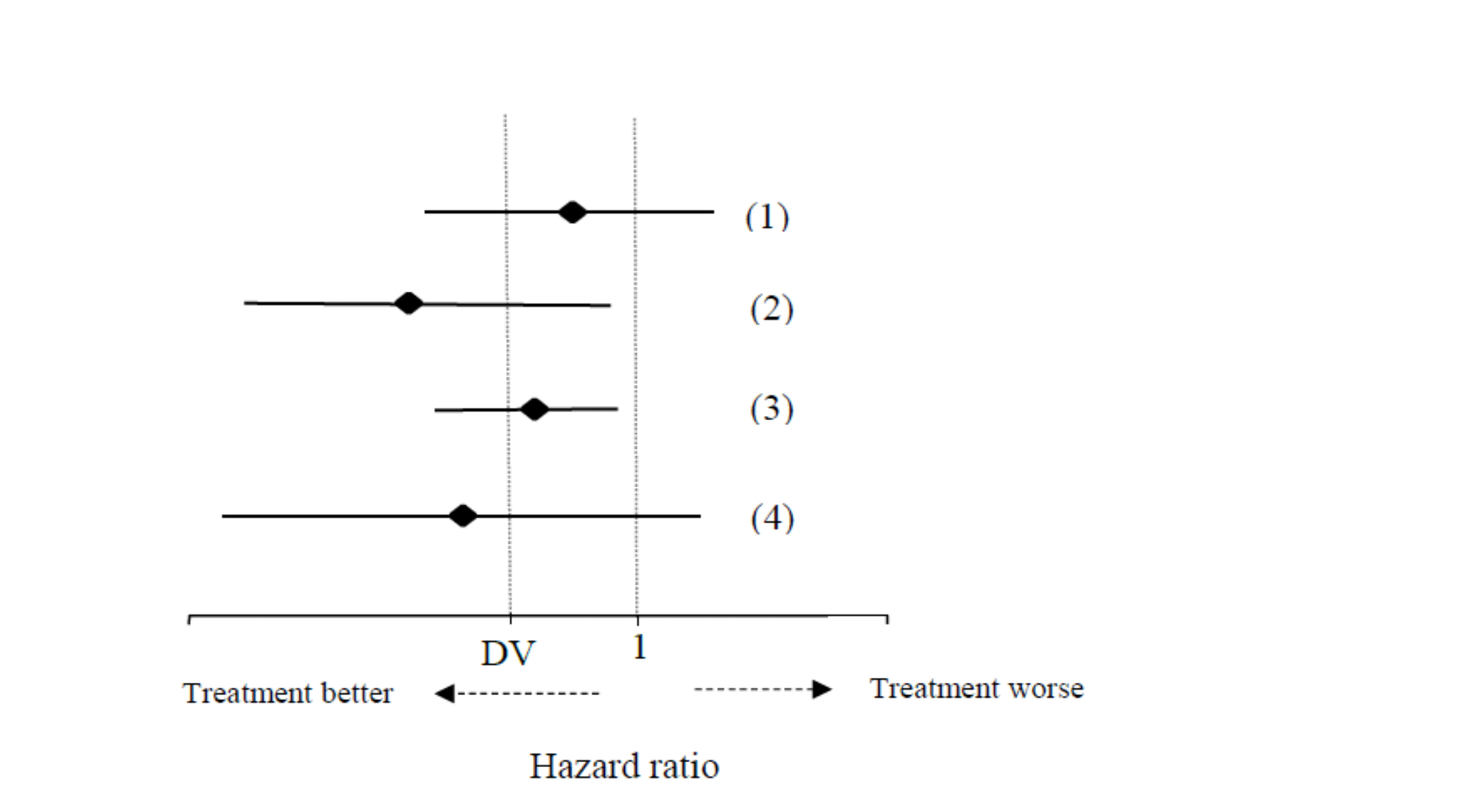}
\par\vspace{0pt}
\end{minipage}
\vspace{0.5 cm}
\caption{Outcomes and decisions for a dual-criterion design. \newline \newline}
\label{dcdecision}
\end{table}

The planning of a dual-criterion trial requires carefully selected
decision criteria and sample size. The dual criterion must be based on
clinical input, which necessitates effective collaborations between
statisticians and clinicians. Understanding the similarities and
differences between a standard and dual-criterion design is
important. Graphs and non-statistical language help to explain the
main features of a dual-criterion design. In addition, data scenarios
with respective decisions may provide further clarification. Finally,
operating characteristics (type-I error and power) are crucial to
justify the design. Details will be discussed in the following
sections.

Since decision making in phase II is complex, a final
decision in inconclusive situations should account for other relevant
information, such as key secondary endpoints, safety, PK/PD,
biomarkers, and subgroups.

In summary, recommendations for selecting a design depend on the
context, which informs the required design inputs. In the comparative
phase II setting, two benchmarks (null value = no effect, decision value =
minimum effect estimate) will often be available, which suggests a
dual-criterion design. In other situations, a standard or precision design
may be more appropriate. Irrespective of the chosen design, a clear
understanding of the development context is crucial, and a consensus
about design inputs and decisions related to possible trial outcomes
is important.

\section{Statistical considerations for dual-criterion designs}
This section provides the key statistical components of dual-criterion designs. These refer to the formulation of the
dual-criterion, the determination of the sample size, and the design's
operating characteristics. 

\subsection{Frequentist and Bayesian dual-criteria}
The frequentist formulation of the null and decision value has been
presented in the previous section. The alternative Bayesian version
requires a sufficiently large probability for a positive treatment
effect (effect exceeds the null value) and an effect estimate that
reaches the decision value.

For the first Bayesian criterion, depending on the required strength
of evidence for a positive effect, 90\%, 95\%, or 97.5\% are common
values for $1-\alpha$. These values are selected in analogy to their
frequentist counterparts (p-value less than one-sided type-I error
level) and will typically lead to type-I errors close to $\alpha$ if
prior information is weak. Importantly, despite the approximate
equality of p-values and Bayesian threshold probabilities, their
meaning is different (Casella and Berger \cite{casella1987}, Berger
and Selke \cite{berger1987}, Wasserstein and Lazar
\cite{Wasserstein2016}).  In addition, the decision value in a
dual-criterion design and the alternative hypothesis in a standard
design have entirely different interpretations. It is therefore
critical to elicit the decision value properly from clinical
considerations. Simply taking the alternative hypothesis from a
standard design as the decision value in a dual-criterion design is
unwarranted.

For example, for a standard time-to-event design with 80\% power at a
one-third hazard reduction (HR=0.667), the implied effect size
threshold (minimum effect estimate for success) is 0.754. Using 0.667
as the decision value in a dual-criterion trial implies a completely
different design. First, the standard design would not fulfill the
clinically relevant criterion (estimate threshold 0.754 instead of
0.667). Second, because the dual-criterion is obviously more
demanding, the resulting power of the study is less compared to the
standard design. For instance, the power at 0.667 is 50\% compared to
80\% for the standard design.

\subsection{Sample size}
After the design inputs $\alpha$, null value (NV), and decision value
(DV) have been set, the sample size can be derived. For normally
distributed data, the minimum sample size is
\begin{eqnarray*}
   n_{min}=  \frac{\sigma^{2}  \times  {z_{\alpha}}^{2}}{(NV-DV)^{2}}    
\end{eqnarray*}
Here, $z_{\alpha}$ is the 100(1-$\alpha$)\%-quantile of the standard
normal distribution, and $\sigma$ is the outcome standard deviation;
for example, $\sigma$ is 2 (under equal randomization) for the
standard normal approximation to time-to-event data.

The $n_{min}$ above is the minimum sample size that implies
statistical significance if the effect estimate equals the decision
value. Importantly, the sample size of a dual-criterion design must be
at least $n_{min}$. It can of course be larger than $n_{min}$ if
operating characteristics are considered unsatisfactory otherwise.

For non-normal data, a grid search over sample sizes may be needed to
determine the minimum sample size. A search over increasing sample
sizes is performed until the sample size $n_{min}$ is found so that
for all sample sizes equal to or larger than $n_{min}$, the
dual-criterion is fulfilled. For an example, see the single-arm trial
design in the next section.

\subsection{Operating characteristics}
Operating characteristics (type-I error, power) are an integral part
of clinical trials designs and should be provided in the study
protocol.

For dual-criterion designs, it is important to understand that the
power at the decision value is (approximately) 50\%: if the true parameter
equals the decision value, there is a roughly equal chance that the
effect estimate lies on either side of the decision value. Notingly,
increasing the sample size will not change the power at the decision
value: for parameters superior to the decision value the power
increases, whereas for inferior values it decreases. This is warranted
because superior values justify a GO, whereas inferior values do
not. Thus, if the true value equals the decision value, a power higher
(lower) than 50\% would bias decisions towards GO (NO-GO), which is
unwarranted. Therefore, having 50\% power at the decision value value
does not mean that the study is underpowered. Such a claim fails to
understand the difference between the decision value of a
dual-criterion design and the alternative parameter of a standard
design.
 
\begin{figure*}[ht]
\centering
\includegraphics[scale=0.6]{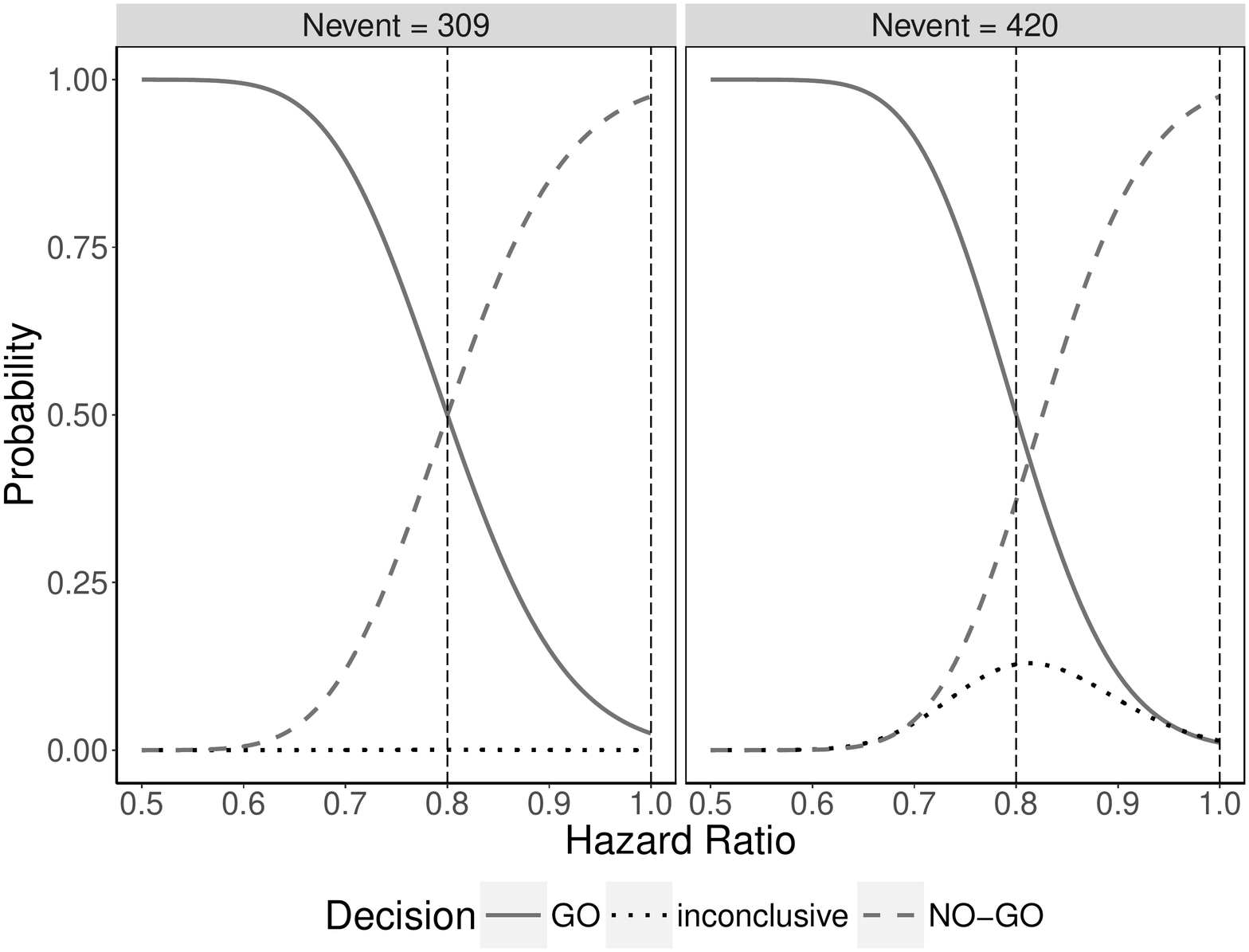}
\caption{Operating characteristics of dual-criterion designs with 309 and 420 number of events.}
\label{DCOC}
\end{figure*}

For the previous example, the minimum number of events ($n_{min}$) is
309. The left panel of Figure \ref{DCOC} shows the operating
characteristics. If power for larger effects than the decision value
is considered unsatisfactory, the sample size should be
increased. However, for larger sample sizes, the effect estimate may
not reach the decision value even if statistical significance holds,
which is the inconclusive case 3 in Table \ref{dcdecision}. Moreover,
larger sample sizes reduce the type-I error below the required
$\alpha$. The right panel of Figure \ref{DCOC} shows the operating
characteristics for 420 events: for effect sizes superior to the
decision value (HR $\leq$ 0.8), the power increases compared to 309
events, whereas for inferior values it decreases.

In addition to the operating characteristics of a dual-criterion
design, data scenarios (effect estimates) with corresponding decisions
may help to clarify decisions obtained from a dual-criterion
design.

\subsection{Final analysis}
The final analysis of a trial with a dual-criterion design does not
differ from other designs. One needs to calculate the respective
frequentist or Bayesian metrics:
\begin{itemize}
\item the p-value (or one-sided confidence interval) and the parameter
  estimate (e.g. the unbiased or maximum-likelihood estimate)
\item the posterior probability that the parameter is better than the
  null-value and the posterior median (or mean). In addition, ploting
  the posterior density and cumulative distribution function may be
  useful.
\end{itemize}
For phase II trials, the ultimate GO or NO-GO decision is based on
these dual-criterion metrics and other possibly relevant information.

\section{Examples}
We now discuss two recent implementations of dual-criterion designs in
phase II. To illustrate the broad applicability of such designs, we
consider a randomized controlled trial with a time-to-event endpoint
and a single-arm trial with a binary endpoint. The implementation will
be frequentist for the first and Bayesian for the second. We also
investigate alternative designs (Rubinstein et al. \cite{Rubinstein2005}, Sargent
et al. \cite{SARGENT2001}), which shows similarities as well as
differences in comparison to the dual-criterion design.

\subsection{A randomized PoC design with time-to-event data}
Example 1 was a randomized, double-blind PoC study of an experimental
drug in combination with standard of care in patients with
non-small-cell lung cancer. Patients were randomized equally to the
experimental drug plus standard of care or to standard of care alone. The primary
endpoint was progression-free survival (PFS), assessed with a log-rank
test and Cox regression with treatment as a covariate.

In addition to the null value of no effect (HR=1), the specification
of the decision value took into account standard of care data, the
competitive landscape, and discussions with clinicians, marketing
experts, and health authorities. A literature review revealed a median
PFS of approximately five months. Based on discussions with experts,
an estimated improvement of at least two months, which corresponds to
an approximate HR of 0.7, was deemed necessary to be clinically
meaningful. Values larger than 0.7 were judged unsatisfactory to
clearly justify further development of the experimental drug.  The
dual-criterion for further development (GO) of the experimental drug
was thus
\begin{enumerate} 
 \item statistical significance: one-sided p-value of log-rank test $\leq$ 0.1
 \item clinical relevance: estimated HR from Cox regression $\leq$ 0.70 
\end{enumerate}
The sample size calculation used approximate normality of the
log-hazard-ratio. From the sample size formula in previous section,
the minimum number of events $n_{min}$ is 52. That is, with at least
52 events, if the estimated HR is 0.7, the one sided p-value is
0.1. To improve the power of the study for effect sizes better than
0.7 (decision value), the number of PFS events was increased to 70. To
reach 70 events, approximately 200 patients (100 per arm) were
expected to be enrolled.

Table \ref{tab::ex1} shows the operating characteristics of the design,
that is, the probability for a GO, NO-GO, or inconclusive
outcome.Design 1 was the actually used design. Since 70 exceeds the
minimal number of events (52), an inconclusive decision (significant
p-value with an estimate inferior to the estimate threshold 0.7) may
occur. However, this probability is less than 0.1 for all hazard ratio
scenarios. The type-I error is 0.032, considerably less than the
required 0.1, and the power is reasonably large for convincing hazard
ratios (0.92 for HR 0.5, 0.74 for HR 0.6).  

The second design considers the case of 52 events, the minimum number
such that statistical significance implies clinical relevance (hazard
ratio estimate = 0.7). As can be seen, inconclusive decisions cannot
arise. Compared to the first design, the type-I error is now the
postulated 0.1, and the power is smaller (compared to design 1) for
hazard ratios superior to the 0.7 threshold. The probability that the
study fails is larger for hazard ratios inferior to 0.7.

For comparison, we now look at three more classical designs with
different choices for type-I error and power at an alternative hazard
ratio of 0.5, referred to as {\em Randomized Screening Designs} by
Rubinstein et al. \cite{Rubinstein2005}. Design 3 is the most
demanding, with one-sided type-I error 0.1 and power 0.9. The number
of events is 55, and statistical significance is achieved if the
hazard ratio estimate is better than 0.708; in the actual trial such
an estimate would have been insufficient for a GO decision. The sample
size and estimate threshold are similar to design 2, so type-I error
and power are similar too. Power is slightly higher due to the larger
number of events and the weaker estimate threshold (0.708 compared to
0.7). However, design 1 differs in two aspects: if the sample size is
larger than $n_{min}$, inconclusive decisions may arise (statistical
significance but clinical irrelevance); and, the type-I error will be
smaller than $\alpha$.

Design 4 shows an example with the same type-I error (0.1) but
decreased power (0.8). The implied required hazard ratio estimate is
now 0.659, which is clearly more aggressive than the explicitly
postulated 0.7 in the dual-criterion design.  

Finally, design 5 uses type-I error 0.2 and power 0.9. This allows for
more GO decisions: the required hazard ratio estimate is now only
0.761, in stark contrast to the clinically relevant value of 0.7. If
an estimated 24\% hazard reduction had been considered sufficient for
a clear GO, a dual-criterion design with decision value 0.76 would
have been an alternative to design 5.

\begin{table}
  \caption{Example 1: operating characteristics for dual-criterion 
    and standard designs; probabilities for GO, NO-GO and
    inconclusive decisions given by hazard ratio estimate
    $\hat{\theta}$. 
\label{tab::ex1}}
  \centering
\begin{tabular}{cccc}
& \multicolumn{3}{c}{1. dual-criterion design: $\alpha$=0.1, DV=0.7, n=70} \\
     true HR &  GO: $\hat{\theta} \le 0.7$ & NO-GO: $\hat{\theta} > 0.736$
                                          & inconclusive \\
0.5   & 0.920 &  0.053 & 0.027  \\
0.6   & 0.740 &  0.196 & 0.064  \\
0.7   & 0.500 &  0.417 & 0.083  \\
0.8   & 0.288 &  0.636 & 0.076  \\
0.9   & 0.147 &  0.800 & 0.054  \\
1.0   & 0.068 & 0.900 & 0.032  \\[2mm]
& \multicolumn{3}{c}{2. dual-criterion design: $\alpha=0.1$, DV=0.7, n=52} \\
      &  GO: $\hat{\theta} \le 0.7$ & NO-GO: $\hat{\theta} > 0.7$
                                           & inconclusive \\
0.5   & 0.887 & 0.113 &  --- \\
0.6   & 0.711 & 0.289 &  --- \\
0.7   & 0.500 & 0.500 &  --- \\
0.8   & 0.315 & 0.685 &  --- \\
0.9   & 0.182 & 0.818 &  --- \\
1.0   & 0.099 & 0.901 &  --- \\[2mm]
& \multicolumn{3}{c}{3. standard design: $\alpha=0.1, \beta=0.1
  (\theta_A=0.5)$, n=55} \\
      &  GO: $\hat{\theta} \le 0.708$ & NO-GO: $\hat{\theta} > 0.708$
                                           & inconclusive \\
0.5   & 0.901 & 0.099 &  --- \\
0.6   & 0.729 & 0.270 &  --- \\
0.7   & 0.516 & 0.484 &  --- \\
0.8   & 0.325 & 0.675 &  --- \\
0.9   & 0.186 & 0.813 &  --- \\
1.0   & 0.100 & 0.900 &  --- \\[2mm]
& \multicolumn{3}{c}{4. standard design: $\alpha=0.1, \beta=0.2
  (\theta_A=0.5)$, n=38} \\
      &  GO: $\hat{\theta} \le 0.659$ & NO-GO: $\hat{\theta} > 0.659$
                                           & inconclusive \\
0.5   & 0.804 & 0.196 &  --- \\
0.6   & 0.615 & 0.385 &  --- \\
0.7   & 0.428 & 0.572 &  --- \\
0.8   & 0.276 & 0.724 &  --- \\
0.9   & 0.169 & 0.831 &  --- \\
1.0   & 0.100 & 0.900 &  --- \\[2mm]
& \multicolumn{3}{c}{5. standard design: $\alpha=0.2, \beta=0.1
  (\theta_A=0.5)$, n=38} \\
      &  GO: $\hat{\theta} \le 0.761 $ & NO-GO: $\hat{\theta} > 0.761$
                                           & inconclusive \\
0.5   & 0.902 & 0.098 &  --- \\
0.6   & 0.768 & 0.232 &  --- \\
0.7   & 0.602 & 0.398 &  --- \\
0.8   & 0.439 & 0.561 &  --- \\
0.9   & 0.303 & 0.697 &  --- \\
1.0   & 0.200 & 0.800 &  --- \\[2mm]
\end{tabular}
\end{table}

\subsection{A single-arm PoC design with binary data}
Example 2 was a single-arm PoC trial of an experimental drug in
Chinese patients with non-small-cell lung cancer. This type of study
is common in early phase Oncology trials, for example in the expansion
phase of a dose-escalation trial. The primary endpoint was objective
response rate (ORR), which quantifies the preliminary efficacy of the
experimental drug. The final analysis, which was Bayesian in this
example, used a binomial sampling and a minimally informative unimodal
Beta prior distribution with mean 0.075, that is, a Beta(0.0811,1)
distribution.

\begin{figure*}[ht]
\centering
\includegraphics[scale=0.5]{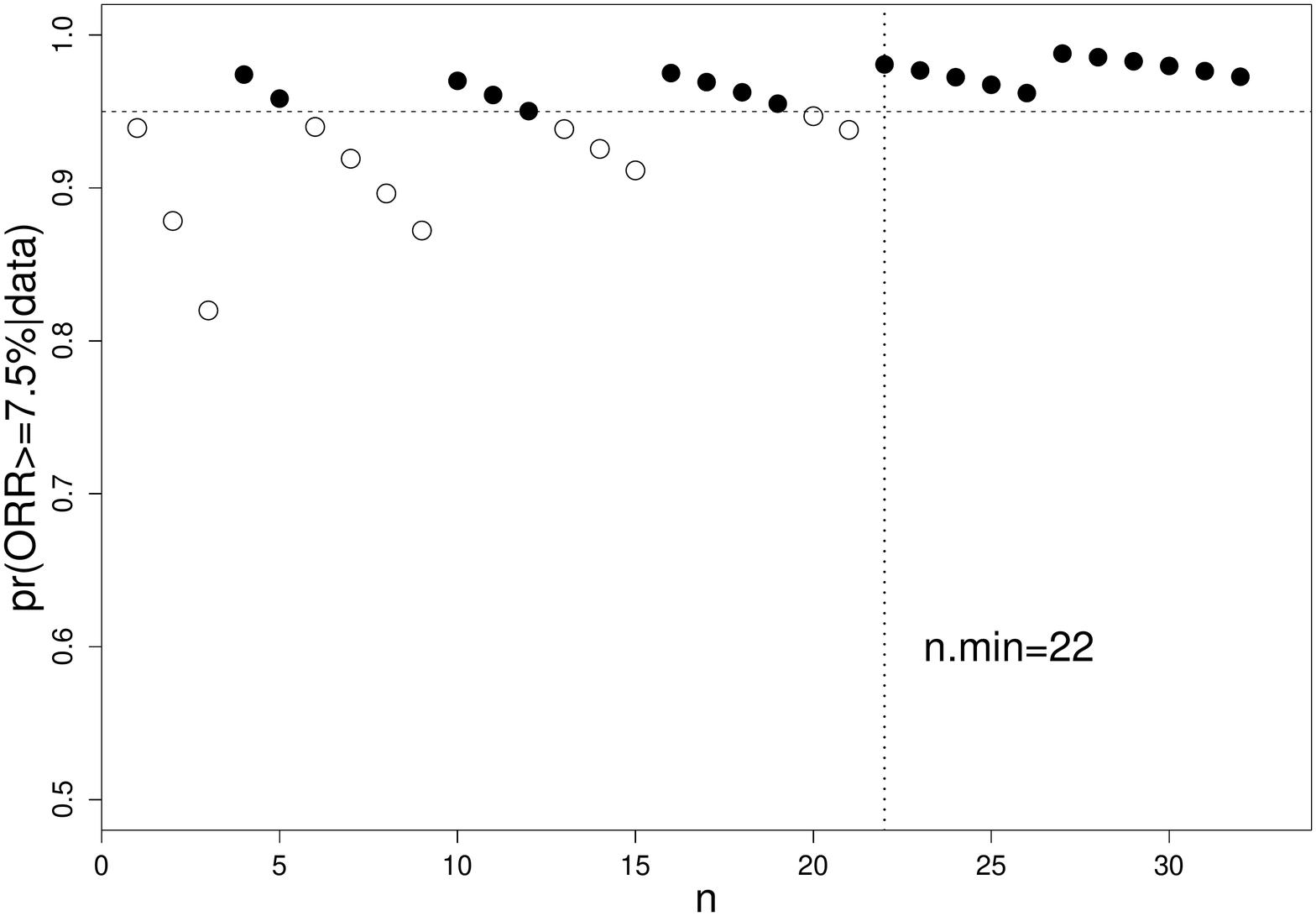}
\caption{Example 2: grid search for minimum sample size.}
\label{Binmin}
\end{figure*}

While the null value is usually clear (no effect) in randomized
trials, selecting it is more challenging in single-arm trials due to
the absence of a comparator. Based on a literature review and clinical
discussions, the null value for the ORR was set to 7.5\%. Moreover, a
minimum improvement of 10\% was considered necessary to justify
further development. The dual-criterion was thus defined as
\begin{enumerate} 
 \item	Bayesian statistical significance:  pr( ORR $\geq$ 7.5\% $|$ data) $\geq$ 0.95
 \item	clinical relevance: posterior median $\geq$ 17.5\%
\end{enumerate}
For this dual-criterion, the minimally required sample size $n_{min}$
is 22;  that is, for $n \geq 22$, clinical relevance (criterion 2) ensures
statistical significance (criterion 1). The final sample size of the
trial was set to 25. Figure ~\ref{Binmin} shows the posterior 
probability that the ORR exceeds 7.5\% as a function of $n$, assuming
that the number of responders is the smallest $r$ such that the
posterior median exceeds the decision value 17.5\%.

For n=25, the minimum number of responders required for a GO is five, with
respective estimate 18.7\% (posterior median) and posterior
probability for a positive effect (ORR $\ge$ 7.5\%) of 0.967. If the
number of responders is less than five, both criteria are missed
(NO-GO): for four responders, the posterior median is 14.8\%
and the probability of a positive effect is 0.895. Thus, in contrast to
example 1, there will be no inconclusive outcome here.

Table \ref{tab::ex2} shows the designs's probabilities of GO or NO-GO
(design 1). The type-I error is 0.036, and the power was considered
reasonably large.

For comparison, we consider a three-outcome design with $H_{0}$: ORR
$\le$ 7.5\%, $H_{1}$: ORR $\ge$ 27.5\%, $\alpha$ = 0.05, $\beta$ = 0.1,
$\eta$ = 0.8 and $\pi$ = 0.9, which requires 27 patients (for details
see Sargent et al. \cite{SARGENT2001}). The decisions are as follows:
NO-GO for zero to three responders, GO for five or more responders,
and inconclusive for four responders.

For comparison to the three-outcome design, assume a larger size
(n=36). Now, to meet the dual-criterion, seven responders are
necessary: for this, the posterior median is 18.5\%, and the
probability of a positive effect is 0.985. An inconclusive outcome
would occur for six responders: the posterior median is 15.8\%
(missing clinical relevance), yet the posterior probability of a
positive effect is 0.954, reaching the 0.95 threshold. The operating
characteristics for this design are also shown in Table \ref{tab::ex2}
(design 2). The comparison to the three-outcome design is
interesting. The type-I error $\alpha$ and type-II error (for a
response rate 0.275) are both small (0.016 and 0.044). In addition,
the probability for NO-GO under the null hypothesis is 0.95, and the
probability for GO is 0.902 for a response rate of 0.275. Both exceed
0.8, as suggested by Sargent et al.  \cite{SARGENT2001}. This shows
that this dual-criterion design is a three-outcome design with
desirable properties.

Finally, design 3 in Table \ref{tab::ex2} shows possible differences
between a dual-criterion and a three-outcome design. First, note that
the operating characteristics reveal small error rates and reasonably
large power, implying a three-outcome design with acceptable
properties. Of special interest is the inconclusive outcome (four
responders in 27 patients): the posterior median is 13.7\% and the
probability of a positive effect (ORR $\ge$ 7.5\%) is 0.869. Thus, the
dual-criterion fails for both (Bayesian) statistical significance and
clinical relevance (NO-GO), whereas the three-outcome design results
in an inconclusive outcome.

\begin{table}
  \caption{Example 2: operating characteristics for dual-criterion 
    and three-outcome designs; probabilities for GO, NO-GO and
    inconclusive decisions given by the number of responders (r).
    \label{tab::ex2}}
  \centering
\begin{tabular}{cccc}
& \multicolumn{3}{c}{1. dual-criterion design: $\alpha$=0.05, DV=0.175, n=25} \\
     true ORR (\%) &  GO: $r \ge 5$ & NO-GO: $r<5$
                                          & inconclusive \\
 7.5 & 0.036 & 0.964 &     --- \\
 12.5 & 0.195 & 0.805 &     --- \\
 17.5 & 0.451 & 0.549 &     --- \\
 22.5 & 0.693 & 0.307 &     ---  \\
 27.5 & 0.858 & 0.142 &     ---  \\[2mm]
& \multicolumn{3}{c}{2. dual-criterion design: $\alpha$=0.05, DV=0.175, n=36} \\
      &  GO: $r \ge 7$ & NO-GO: $r \le 5$
                                          & inconclusive: $r=6$ \\
 7.5 & 0.016 & 0.950 & 0.033 \\
 12.5 & 0.156 & 0.709 & 0.135 \\
 17.5 & 0.446 & 0.380 & 0.174 \\
 22.5 & 0.731 & 0.149 & 0.121 \\
 27.5 & 0.902 & 0.044 & 0.054 \\[2mm]
& \multicolumn{3}{c}{3. three-outcome design: n=27} \\
      &  GO: $r \ge 5$ & NO-GO: $r \le 3$
                                          & inconclusive: $r=4$ \\
 7.5 & 0.048 & 0.860 & 0.092 \\
 12.5 & 0.243 & 0.558 & 0.199 \\
 17.5 & 0.523 & 0.280 & 0.197 \\
 22.5 & 0.759 & 0.113 & 0.128 \\
 27.5 & 0.901 & 0.038 & 0.062 \\[2mm]
\end{tabular}
\end{table}

The computations for the two examples of this section were done with R 3.2 \cite{r32}. Code is available from the main author upon request.

\section{Discussion}
The development plan of each experimental drug is unique, for which phase
II trials are an important screening or early examination step. A good
phase II design should be based on clinical needs, already available
information about the drug, and the competitive landscape. In
addition, it must have acceptable error rates for well-defined success
criteria. For comparative efficacy trials, standard designs aim for a
statistically significant p-value under type-I error control and
sufficiently large power. However, statistical significance only
guarantees sufficient evidence that the drug has an effect. It does
not guarantee clinical relevance.

We have discussed a dual success criterion that goes beyond a
single null-metric, a p-value (Wasserstein and Lazar
\cite{Wasserstein2016}) or posterior threshold probability. The criterion
complements statistical significance by a sufficiently large effect
estimate. The latter is easily interpretable, ensuring clinical
relevance if success is declared. Whether a frequentist or Bayesian
design is chosen is rather unimportant in the absence of prior
information. Otherwise, a Bayesian implementation is
preferred. 

There are similarities for dual-criterion and standard designs. Type-I
error $\alpha$ and power $(1-\beta)$ are important for both. For
standard designs, they define the design, leading to an implicit
success threshold for the effect estimate. To guarantee clinical
relevance (a sufficiently large effect estimate), the effect threshold
can be changed by changing $\alpha$ and $\beta$. Alternatively, and
more directly, a dual-criterion can be used. This, however, has no
explicit power requirements, and power should therefore
be assessed. Irrespective of the chosen design, the three main metrics
(type-I error, power, estimation threshold for success) must be understood to
ensure a sensible balance of statistical and clinical requirements. 

Beyond these similarities, differences between the two designs should
be kept in mind. First, standard designs always result in success
(statistical significance) or failure. Dual-criterion designs may lead
to a statistically significant but clinically irrelevant result. As
we have shown, such inconclusive outcomes are possible if the sample
size is larger than the minimum sample size that implies statistical
significance if the estimate reaches the clinical success threshold.
Second, for standard designs, increasing the sample size increases the
power for effects better than the null. For dual-criterion designs,
power is only increased for values superior to the decision value
since inferior values are clinically irrelevant.

The specification of a dual-criterion should be tailored to the
development plan. While conceptually simple, its implementation
requires care. We have experienced two related challenges. First, the
elicitation of the decision value, the threshold for the effect
estimate, is sometimes considered challenging. However, if carefully
done, elicitation of the decision value is often easier for clinicians
compared to an alternative hypothesis with respective power. Second,
the fact that study power is 50\% at the decision value has let some
reviewers conclude that the study is underpowered. This concern can be
alleviated in the protocol by precisely defining the dual-criterion,
clarifying that the decision value differs from the alternative
hypothesis in a standard design, and providing the power for effect
parameters considerably better than the decision value.

We have only discussed the dual-criterion design for two simple phase
II examples.  A dual-criterion can of course be applied in other
settings too because requirements of statistical significance and a
reasonably large effect estimate are generally applicable. 
Examples include non-inferiority (Neuenschwander et
al.\cite{neuenschwander2011}), bridging, and dose-finding studies, as
well as multiple endpoints. In addition, dual-criterion designs may
include interim analyses for futility for which we often use
probability of success for interim decisions (Gsponer et
al. \cite{gsponer2014}). 

Finally, our experience so far has been confined to nonconfirmatory
trials. In the confirmatory setting, the size of effect estimates
matters but is usually not part of the success criterion.  However, a
formal criterion may help to avoid negative surprises, in particular
for large trials: rather than questioning the relevance of a
significant phase III trial with an unconvincing effect estimate,
formally introducing the minimum effect estimate in the success
criterion may be a better option.

\begin{acks}
  We would like to thank Santosh Sutradhar, Haige Shen, and Titing Yi
  for their contributions, as well as Michael Branson, Nathalie
  Fretault, Daniel Lorand, Doug Robinson, and Simon Wandel for
  valuable discussions and support. In addition, we also thank the two
  referees and the associate editor for their thoughtful comments,
  which greatly helped to improve the manuscript.
\end{acks}



\bibliographystyle{SageV}
\bibliography{dcreference}

\end{document}